\documentclass[
    ,final            
  ]
  {aipproc}

\layoutstyle{6x9}

\begin{document}

\title{EXPLORE/OC:  A Search for Planetary Transits in the Field of the Southern Open Cluster NGC 6208}

\author{Brian L. Lee}{
  address={Dept. of Astronomy \& Astrophysics, Univ. of Toronto, 60 St. George Street, Toronto, Ontario M5S 3H8 Canada}
}

\author{Kaspar von Braun}{
  address={Department of Terrestrial Magnetism, 5241 Broad Branch Road NW, Washington, DC 20015 USA}
}

\author{Gabriela Mall\'{e}n-Ornelas}{
  address={Harvard-Smithsonian Center for Astrophysics, 60 Garden Street, Cambridge, MA 02138 USA}
}

\author{H. K. C. Yee}{
  address={Dept. of Astronomy \& Astrophysics, Univ. of Toronto, 60 St. George Street, Toronto, Ontario M5S 3H8 Canada}
}

\author{Sara Seager}{
  address={Department of Terrestrial Magnetism, 5241 Broad Branch Road NW, Washington, DC 20015 USA}
}

\author{Michael D. Gladders}{
  address={Carnegie Observatories, 813 Santa Barbara Street, Pasadena, California 91101 USA}
}

\begin{abstract}
%
%
The EXPLORE Project expanded in 2003 to include a campaign to monitor rich southern open clusters for transits of extrasolar planets (EXPLORE/OC).  In May and June 2003, we acquired precise, high-cadence photometry of the second open cluster in our campaign, NGC 6208.  Here, we present preliminary results from our $I$-band survey of over 60000 stars in the field of NGC 6208, around 5000 of which were monitored with photometric precision better than 1\%.

\end{abstract}

\maketitle


\section{Introduction}

The EXPLORE collaboration has previously monitored Galactic plane fields for evidence of transiting close-in extrasolar giant planets
\cite{Mallen:2003}.
Using the experience built up in the deep Galactic plane searches, we are now conducting a complementary study of open clusters.

While the potential number of monitorable stars in the Galactic plane fields is generally higher than in open clusters, an open cluster offers a large sample of stars sharing a common age, metallicity, distance, and reddening.  
By trading off the number of stars that can be monitored, we thus gain control over the properties of our stellar sample (although, since open clusters are concentrated in the Galactic plane, many of the stars will be from the field and not part of the cluster sample;  also, we may see differential reddening across the field).  
In surveying a specific cluster, we accumulate statistics for stars sharing a common set of relatively easily measured properties, and we can target a specific range of spectral types by adjusting the exposure time.
Our campaign will be useful in order to constrain the occurrence of planets as a function of age and metallicity, and will also allow us to compare the frequency of planets in a cluster environment to that in the Galactic field.

NGC 6208 (distance=1.10kpc, E(B-V)=0.35, [Fe/H]=0.00 \cite{Twarog:1997}; age=1.00 Gyr \cite{Piatti:1995})
was chosen out of a list of potential targets because of its observability, its high star count, and a distance and reddening such that with five-minute exposures, cluster G and K dwarfs would typically be monitored with better than 1\% photometry in $I$.  
We conduct our survey in the $I$-band (rather than $V$ or $R$) to minimize the effects of limb-darkening on the slope of a transit's ingress and egress, allowing for better discrimination of good transit candidates from other low-amplitude variations (such as blends and grazing binaries) \cite{Seager:2003}.

Here, we present some of our observations, data reduction techniques, and preliminary results from our survey of NGC 6208.
Preliminary results of the study of the first target in the EXPLORE/OC campaign, NGC 2660, are presented in a companion paper \cite{vonBraun:2004} in this volume.

\section{Observations}

From May 16 to June 19, 2003, we observed NGC 6208 
($l=333.7^\circ$, $b=-5.8^\circ$) using the Swope 1-m telescope at Las Campanas Observatory, Chile.  We obtained 1730 $I$-band monitoring frames over $\sim 22$ clear nights out of a run one-third interrupted by poor winter weather (the effect of this loss on transit detection is illustrated in Figure \ref{fig:pvis}).  When monitoring NGC 6208, we took successive 300 sec. $I$-band exposures of a 15'$\times$23' field centred on the cluster, imaged onto 2048$\times$3350 CCD pixels, with 130 seconds readout.  On a photometric night, we also obtained a handful of $BVR$ frames for photometric calibration and construction of colour-magnitude diagrams, and $BVRI$ frames of a nearby off-cluster comparison field centred at ($l=334.7^\circ$, $b=-5.8^\circ$).  We have collected low-resolution ($R$ $\approx$ 2000) spectra of selected variable and selected typical stars from our field ($\sim 50$ in all) for MK classification;  these data were taken on July 27 and 28, 2003, using the B\&C spectrograph on the Magellan Baade 6.5-m telescope at Las Campanas, using the 1200 lines/mm grating blazed at 4000\AA, providing a spectral coverage of 3607\AA-5239\AA~and a dispersion of 0.8\AA/pixel.

\section{Analysis and Results} 

\begin{figure}
  \includegraphics[height=.28\textheight]{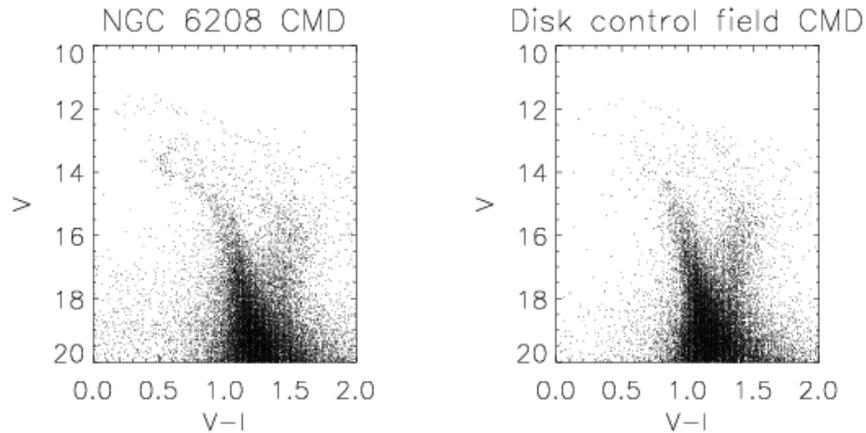}
\caption{Colour-magnitude diagrams of NGC 6208 ($l=333.7^\circ$, $b=-5.8^\circ$) and an off-cluster comparison field ($l=334.7^\circ$, $b=-5.8^\circ$).  The bright end of the cluster main sequence appears clearly on top of the Galactic field population in the NGC 6208 colour-magnitude diagram.}
\label{fig:cmd}
\end{figure}

The $BVRI$ data were used to construct colour magnitude diagrams of the NGC 6208 field and the nearby Galactic field (see Figure \ref{fig:cmd}).  The cluster main sequence is clearly distinguishable from the Galactic field population for $I<13.5$.  There, the star count for the cluster field is $21$\% above the count for the comparison field.  The percentage of Galactic field stars per magnitude bin grows towards fainter magnitudes.

\begin{figure}
  \includegraphics[height=.563\textheight]{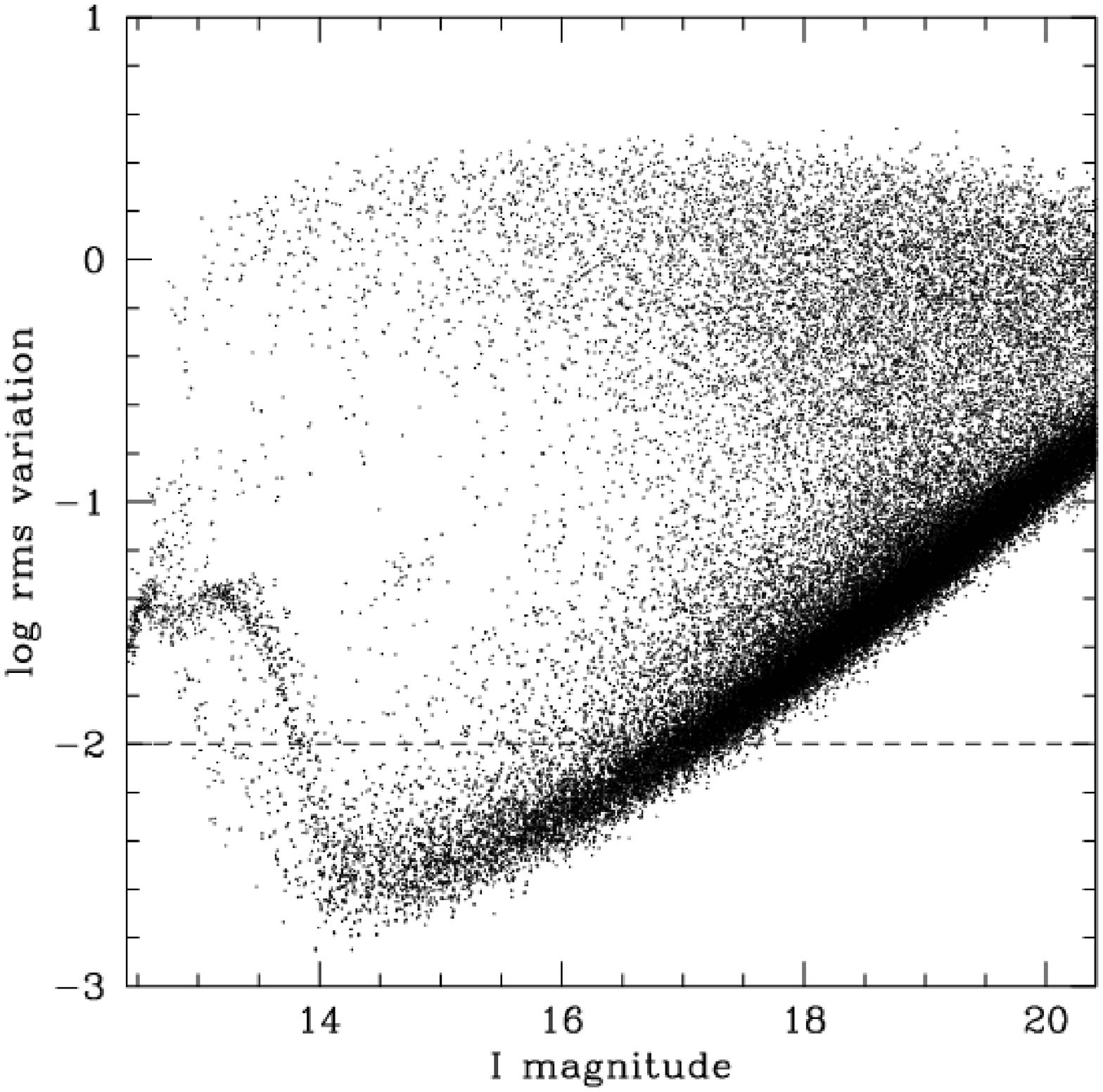}
  \includegraphics*[height=.75\textheight]{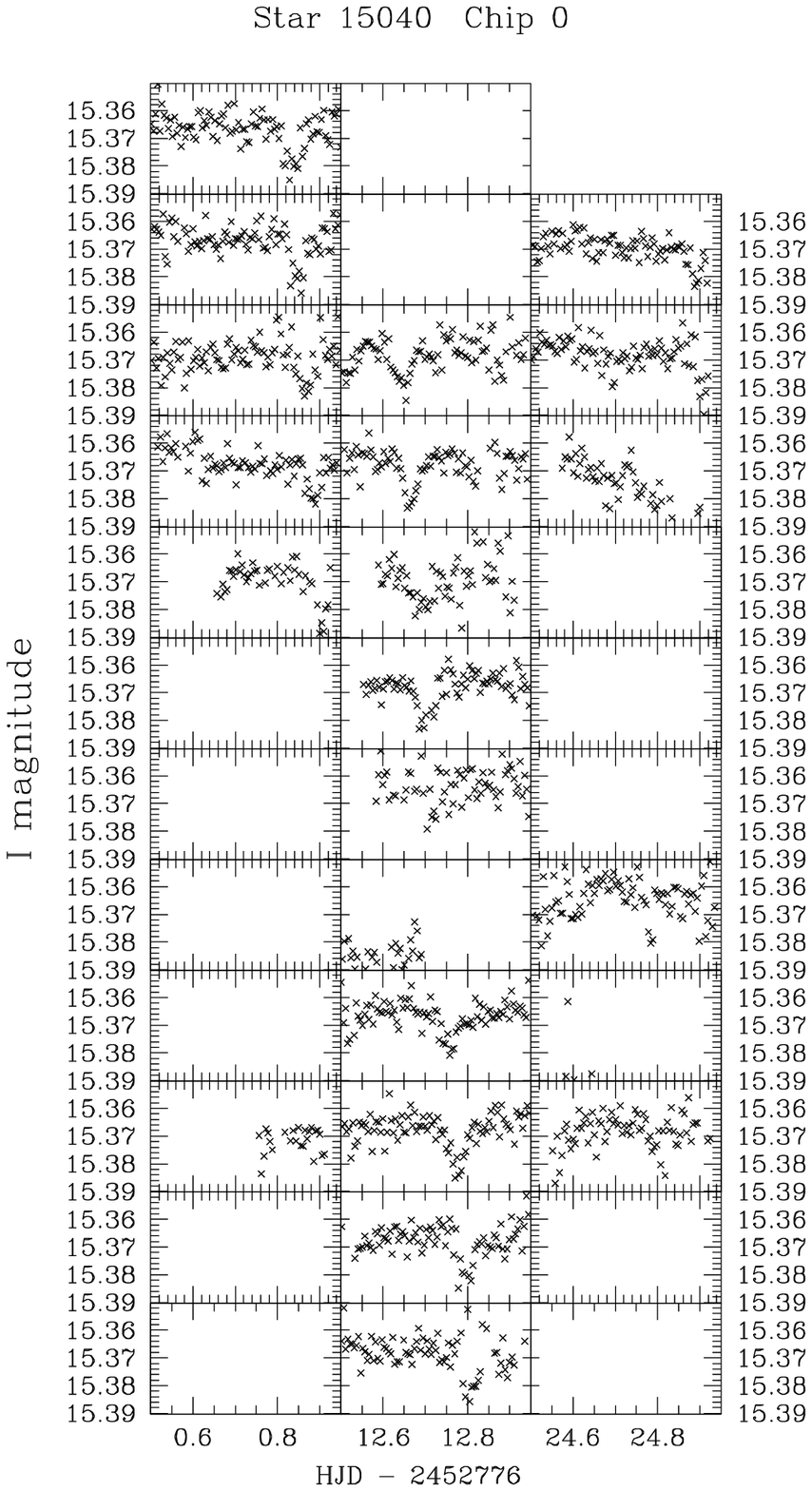}
\caption{\textbf{Left panel:}  Precision of one night of $I$-band photometric monitoring of the field of NGC 6208.  Approximately 5000 stars show RMS variation less than 1\% (indicated by the dashed line) on this night.  \textbf{Right panel:}  Light curve of an $I=15.4$ star in the field, with a periodic transit-like photometric signal at the $\sim$2\% level.  The light curve from the first night (HJD=2452776) of the run is in the bottom left;  light curves from successive nights are shown progressing upwards to the top of the page (then wrapping to the bottom of the next column).  No monitoring was done on nights 1, 2, 4-7, 23-26, 31, and 32.}
\label{fig:rmslc}
\end{figure}

The monitoring data were analyzed using the existing EXPLORE photometric pipeline
\cite{Mallen:2003}.
This fast pipeline allowed us to make light curves of a given night's data with one day's turnover, and hence to begin looking for transit-like features while still at the telescope.  Figure \ref{fig:rmslc} shows the root mean square variation in the $I$ magnitude over one night's monitoring, and demonstrates that we achieve 1\% precision from $I \approx 17$ up to saturation of the detector near $I \approx 14.2$.

\begin{figure}
  \rotatebox{-90}{\includegraphics[height=.25\textheight]{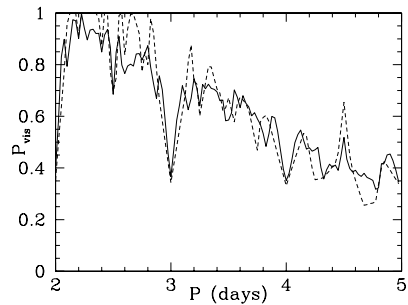}}
\caption{Probability $P_{vis}$ of detecting an existing transiting planet, assuming the signal is large enough to detect, as a function of orbital period $P$ in days, and averaged over all phases.  A detection requires two transits to occur during the nights of an observing run.  The solid line shows the total probability for our actual NGC 6208 run;  the probability is similar to that achieved for a hypothetical perfect run of only 18 consecutive 10.8 hour nights, shown by the dashed line.  Transits with integer periods are statistically difficult to detect because for approximately half the phases, the transits will always occur during the day.  The mean $P_{vis}$ for 2-5 day periods for our run was 62\%.}
\label{fig:pvis}
\end{figure}

We examined, by eye, all of the best light curves (unphased photometric precision better than $\sim$ 1.5\%, $\sim$ 15000 light curves).
To flag a light curve as a candidate transit, we required the detection of at least two low-amplitude eclipse-like features.  Our window function for detecting two transits is shown in Figure \ref{fig:pvis}.

In our data, we found one light curve featuring a transit-like signal (see Figure \ref{fig:rmslc}).  Preliminary spectroscopic data indicate this star's spectral type is late G;  however, because the ingresses and egresses of the eclipses are relatively shallow, it is unlikely to be a planet \cite{Seager:2003}.  While this star may not harbour a transiting planet, it is one example of our ability to rapidly detect variability at the 1\% level in our open cluster surveys.

\bibliographystyle{aipproc}

\end{document}